\documentclass[12pt]{article}
\usepackage{epsfig}
\usepackage{bm}
\usepackage{graphicx}
\usepackage{cite}
\usepackage{amsfonts}
\usepackage{amssymb}
\usepackage{latexsym}
\def\ee{\end{equation}}
\def\ba{\begin{eqnarray}}
\def\ea{\end{eqnarray}}

\def\bq{\begin{quote}}
\def\eq{\end{quote}}

 at 10truept

\newcommand{\beq}{\begin{equation}}
\newcommand{\eeq}{\end{equation}}
\newcommand{\beqa}{\begin{eqnarray}}
\newcommand{\eeqa}{\end{eqnarray}}
\newcommand{\bea}{\begin{eqnarray}}
\newcommand{\eea}{\end{eqnarray}}
\newcommand{\p}{\partial}
\newcommand{\al}{\alpha}
 \newcommand{\be}{\beta}

\newcommand{\bra}{\left<}
\newcommand{\ket}{\right>}
\newcommand{\lb}{\left|}
\newcommand{\rb}{\right|}
\newcommand{\vect}[1]{\bm{\mathrm{{#1}}}}

\def\ltap{\ \raise.3ex\hbox{$<$\kern-.75em\lower1ex\hbox{$\sim$}}\ }
\def\gtap{\ \raise.3ex\hbox{$>$\kern-.75em\lower1ex\hbox{$\sim$}}\ }
\def\gl{\ \raise.5ex\hbox{$>$}\kern-.8em\lower.5ex\hbox{$<$}\ }
\def\roughly#1{\raise.3ex\hbox{$#1$\kern-.75em\lower1ex\hbox{$\sim$}}}

\begin{document}

\thispagestyle{empty}
\begin{flushright}
CERN-PH-TH/2010-111
\end{flushright}

\vskip2cm
\begin{center}
{\Large{\bf On the Effective Equation of State of Dark Energy }}
\vskip2cm {\large Martin S. Sloth\footnote{\tt sloth@cern.ch}}\\

\vspace{.5cm}

\vskip 0.1in

{\em CERN, Physics Department, Theory Unit, CH-1211 Geneva 23, Switzerland}\\

\vskip 0.1in
\vskip 0.1in
\vskip .25in
{\bf Abstract}
\end{center}
In an effective field theory model with an ultraviolet momentum cutoff, there is a relation between the effective equation of state of dark energy and the ultraviolet cutoff scale. It implies that a measure of the equation of state of dark energy different from minus one, $\omega \neq -1$, does not rule out vacuum energy as dark energy. It also indicates an interesting possibility that precise measurements of the infrared properties of dark energy can be used to probe the ultraviolet cutoff scale of effective quantum field theory coupled to gravity. In a toy model with a vacuum energy dominated universe with a Planck scale cutoff, the dark energy effective equation of state is $w_{eff} \approx -0.96$. 


\vskip 2.0 in
\begin{center}
\noindent
{\em \small This essay received an "honorable mention" in the 2010 Essay Competition\\ of the Gravity Research Foundation.}\\
\end{center}
\vfill \setcounter{page}{0} \setcounter{footnote}{0}
\newpage

\setcounter{equation}{0} \setcounter{footnote}{0}

\section{Introduction}

Dark energy is required in order to explain the current acceleration of the Universe \cite{Riess:1998cb,Astier:2005qq,WoodVasey:2007jb,Spergel:2006hy,Tegmark:2006az}. To be consistent with current observations about $70$\% of the energy density of the universe must be dark energy. On the other hand we have a very poor theoretical understanding of the origin of dark energy. It is often believed that a better understanding of dark energy will be accompanied with a radical change in our understanding of quantum field theory coupled to gravity. 

A precise measure of the equation of state parameter, $w$, of dark energy will help us to discriminate between different theoretical models, and has generated a significant interest on the observational side. A detection of $w \neq -1$ is typically believed to be a falsification of vacuum energy driven late time acceleration, and a verification of dynamical dark energy, such as quintessence \cite{Wetterich:1987fm,Peebles:1987ek,Ratra:1987rm,Caldwell:1997ii}.

There might also be an other possibility. Even if the universe is dominated by vacuum energy, the back-reaction of ultraviolet quantum fluctuations can lead to a small deviation from $w =-1$ if the fundamental cutoff on physical momenta is close to the Planck scale. This would lead to the very exciting possibility that we could constrain the value of the ultraviolet cutoff scale of effective quantum field theory coupled to gravity by observing the infrared properties of dark energy.

Since a cutoff on physical momentum violates general covariance, we do not expect energy to be conserved for the quantum modes below the cutoff. The cutoff will therefore have to be associated with a source term for the fluctuations, in order to account for modes redshifting across the new physics hypersurface defined by the cutoff \cite{KeskiVakkuri:2003vj}. In this way, the modes below the cutoff are sourced by the unknown physics above the cutoff.

We will present a simple toy model were there are only two contributions to the energy density of the universe. The dark energy driving the expansion of the background, and the energy density of quantum fluctuations. In order to have overall energy conservation the dark energy component will act as a reservoir for ultraviolet quantum modes. As the ultraviolet quantum modes are continuously generated, the reservoir is slowly emptied, and as a result the effective equation of state of dark energy will deviate slightly from $w=-1$. In the phenomenological picture that emerges, the nature of dark energy is deeply connected to the ultraviolet physics. 

In fact, we will see that if the accelerated expansion of the universe is driven by a dark energy component with equation of state, $w=-1$, then the back-reaction of the ultraviolet quantum modes of a scalar field, will yield an effective equation of state $w_{eff}=-0.96$.

\section{Vacuum energy}

Let us consider for simplicity a massless test scalar field in a quasi de Sitter speace-time. Since we are interested only in ultraviolet effects, the deviations from the massless approximation in the final result will be small as long as the mass is much less than the effective ultraviolet cutoff. In conformal time, the metric is 
\beq	
ds^2 = a^2(t)(-d\eta^2+ d\vect{x}^2)~,
\eeq
where $a(\eta)$ is the scale factor . The Lagrangian of the test field in this metric is given by $\mathcal{L}= (1/2) a^2( \phi'^2+(\nabla \phi)^2)$

The energy density in the vacuum state can be computed from the expectation value of the Hamiltonian. In Fourrier space the Hamiltonian becomes 
\beq\label{H2}
H =\frac{1}{2} \int d^3 k\left[ \frac{1}{a^2} \pi_{\vect{k}}\pi_{\vect{k}}^{\dagger}+a^2(k^2+a^2m^2)\phi_{\vect{k}}\phi_{\vect{k}}\right].
\eeq
where the canonically conjugate field is given by $\pi =a^2\phi'$.

One can quantize the Hamiltonian following the usual prescription in the Heisenberg picture (se e.g \cite{Polarski:1995jg}), by defining the time dependent oscillators 
\bea
\hat a_{\vect{k}}(\eta) &=&\frac{1}{\sqrt{2}}\left(a\sqrt{k}\hat\phi_{\vect{k}}(\eta)+i\frac{1}{a\sqrt{k}}\hat\pi_{\vect{k}}(\eta)\right)\nonumber\\
\hat a^{\dagger}_{\vect{k}}(\eta)&=& \frac{1}{\sqrt{2}}\left(a\sqrt{k}\hat\phi_{-\vect{k}}(\eta)-i\frac{1}{a\sqrt{k}}\hat\pi_{-\vect{k}}(\eta)\right)~,
\eea
which obey the usual commutator relations  $[\hat a_{\vect{k}}(\eta),\hat a^{\dagger}_{\vect{k}'}(\eta)]=\delta^{(3)}(\vect{k}-\vect{k}')$.

The vacuum state at any given time, $\lb 0;\eta \ket$, is defined as the stated annihilated by the lowering operator $\hat a_{\vect{k}}(\eta)\lb 0;\eta \ket=0$. From the normalization of the delta-function in a finite box $\delta^{(3)}(0)=(L/2\pi)^3=V/(a^4(2\pi)^3)$, in the limit of the box size, $L$, going to infinity, we find that the vacuum energy density of the field becomes
\beq
\rho =  \frac{\bra \eta;0 \rb\hat H \lb 0;\eta \ket  }{V} = \frac{1}{4\pi^2 a^4}\int_0^{a\Lambda} dk k^2 k = \frac{1}{16\pi^2}\Lambda^4~.
\eeq
This constant contribution to the energy density can be absorbed in a renormalization of the bare cosmological constant, and we usually do not ascribe any physical significance to it on its own.

\section{Particle production}

Due to particle production in the expanding background, the field will not remain in the initial vacuum state. This is easily seen by making a Bogolubov transformation of the oscillators, in terms of the oscillators at some fixed initial time $\eta_0$,
\bea
\hat a_{\vect{k}}(\eta)&=& \al_{\vect{k}}(\eta)\hat a_{\vect{k}}(\eta_0)+\be_{\vect{k}}(\eta)\hat a^{\dagger}_{-\vect{k}}(\eta_0)\nonumber\\
\hat a^{\dagger}_{\vect{k}}(\eta)&=& \al^*_{\vect{k}}(\eta) \hat a^{\dagger}_{-\vect{k}}(\eta)    +\be^*_{\vect{k}}(\eta)  \hat a_{\vect{k}}(\eta)
\eea
where $\al_{\vect{k}}(\eta)$ and $\be_{\vect{k}}(\eta)$ are the Bogolubov coefficients, which describes  the mixing of raising and lowering operators as time passes. 

Expanding the field and its conjugate in terms of mode functions $f_k$ and $g_k$,
\bea
\phi_{\vect{k}}(\eta) &=& f_k(\eta)\hat a_{\vect{k}}(\eta_0)+f^*_k(\eta)\hat a^{\dagger}_{-\vect{k}}(\eta_0)~,\nonumber\\
\pi_{\vect{k}}(\eta) &=&-i( g_k(\eta)\hat a_{\vect{k}}(\eta_0)-g^*_k(\eta)\hat a^{\dagger}_{-\vect{k}}(\eta_0))~,
\eea
one finds that the mode functions are related to the Bogolubov coefficients,
\bea
f_k(\eta) &=& \frac{1}{\sqrt{2k}}(\al_k(\eta)+\be^*_k(\eta)),\nonumber\\
g_k(\eta) &=&\sqrt{\frac{k}{2}}(\al_k(\eta)-\be^*_k(\eta))
\eea

If we assume that the field starts out in the vacuum initially at $\eta=\eta_0$, then the expectation value of the Hamiltonian in this state at any later time becomes
\beq\label{H2}
\bra \eta_0;0 \rb\hat H \lb 0;\eta_0 \ket = \delta^{(3)}(0)\int d^3 k \left[k\left(|\be_k(\eta)|^2+\frac{1}{2}\right)\right]~.
\eeq

The mode functions of a massless scalar field in de Sitter space are,
\beq
f_k(\eta) =\frac{1}{a\sqrt{2k}}e^{-ik\eta}\left(1-\frac{i}{k\eta}\right)~,\qquad g_k(\eta)=a\sqrt{\frac{k}{2}}e^{-ik\eta}~,
\eeq  
which implies that
\beq
\al_k(\eta)=\frac{1}{2}\left(2-\frac{i}{k\eta}\right)e^{-ik\eta}~,\qquad \be_k(\eta)=-\frac{1}{2}\frac{i}{k\eta}e^{-ik\eta}~.
\eeq
Inserting these expressions back into eq.(\ref{H2}), and subtracting the vacuum energy calculated in the previous section, we find that the renormalized energy density becomes
\beq\label{rhoren}
\rho_{re} \approx \frac{1}{2\pi^2a^4}\int_0^{a\Lambda} dk k^3 |\beta_k(\eta)|^2 = \frac{1}{16\pi^2}H^2\Lambda^2.
\eeq
In the leading approximation $H$ is constant, but in any realistic universe with matter in it, $H$ will slowly decrease and $H^2\Lambda^2$ cannot be cancelled by a cosmological constant term.

The energy density calculated from the Hamiltonian defined in eq.(\ref{H2}),
 agrees with what we would obtain from the expectation value of the energy momentum tensor $T_{\mu\nu} = \p_{\mu}\phi\p_{\nu}\phi-g_{\mu\nu}\mathcal{L}$, when comparing to the ideal fluid form $T^{\mu}_{\nu}=diag.(-\rho,p,p,p)$. We could of course calculate the pressure in the same way. Then we would obtain 
\beq
\hat p =\frac{1}{2V} \int d^3 k\left[ \frac{1}{a^2}\hat \pi_{\vect{k}} \hat\pi_{\vect{k}}^{\dagger}-\frac{1}{3}a^2k^2 \hat\phi_{\vect{k}}\hat\phi_{\vect{k}}\right]
\eeq
Inserting the oscillators, we find that the expectation value of the renormalized pressure becomes
\beq
p_{re} \approx \frac{1}{2\pi^2}\int_0^{a\Lambda} dk k^3 \left[|\beta_k(\eta)|^2-\frac{2}{3}(\al_k\be_k+\al_k^*\be_k^*) \right]= -\frac{1}{3}\frac{1}{16\pi^2}H^2\Lambda^2~,
\eeq
where a term of order $\Lambda^4$ related to the pressure of the vacuum energy, has been subtracted from the bare pressure.

The method of renormalizing the energy-momentum tensor by subtracting its expectation value in the initial Bunch-Davies vacuum is a standard approach. It has been applied to the calculation of back-reaction in the trans-Planckian literature, where it was shown that in the non-standard case of large $|\beta_k|$, there is a problem with large back-reaction unless the cutoff is very low \cite{Tanaka:2000jw, Starobinsky:2001kn}. However, even in the standard case of a small $\beta_k$ like in eq.~(\ref{rhoren}) above, there is a small but still important back-reaction effect, if the cutoff is sufficiently close to the Planck scale.

\section{Back-reaction and the effective equation of state}

From eq.(\ref{rhoren}) we see that since $H$ is assumed to be almost constant, the ultraviolet quantum fluctuations of the scalar field seems to contribute a nearly constant energy density $\rho_{re}\sim const.$ Naively, there might seem to be a contradiction between the continuity equation
\beq\label{conteq}
\dot\rho+3H(\rho+p)=0,
\eeq
which implies that a constant energy density is always associated with a negative but equal pressure, and the equation of state of our produced particles $p_{re}=-\frac{1}{3}\rho_{re}$. However, one should note that the energy conservation equation is violated by the fixed cutoff, since the cutoff acts as an energy reservoir from which energy is pumped into the system, when ultraviolet modes redshifts across it.

This effect can be taken into account by the use of a source term. For an almost constant energy density, ${T^{\mu 0}}_{;\mu}=0$ would imply that $\rho \approx -p$. But obviously ${T^{\mu 0}}_{;\mu}=0$ is not valid for the ultraviolet modes, since modes red-shift  across the cutoff and energy leaks into the system that is no longer closed. One therefore needs to include a source term on the right-hand-side modifying the continuity equation, such that eq.(\ref{conteq}) for the ultraviolet modes becomes $\dot\rho_{re}+3H(\rho_{re}+p_{re})=Q_{re}$, 
where $Q_{re}$ is the energy transfer. In order for the equation to be satisfied with $\dot \rho_{re} = 0$ and $p_{re}=-\frac{1}{3}\rho_{re}$, we must have $Q_{re}=2H\rho_{re}$.

The dark energy component, which drives the (quasi) de Sitter expansion, is the only other component of energy density. So, to have overall energy conservation, we must compensate with an equal but opposite source term in the dark energy fluid. In this picture, vacuum energy acts as a reservoir for ultraviolet quantum modes. As the ultraviolet quantum modes are continuously generated, the reservoir is slowly emptied. 

If we denote the energy density of dark energy by $\rho_{\Lambda}$, then for the dark energy fluid we have $\dot\rho_{\Lambda}+3H(\rho_{\Lambda}+p_{\Lambda})=Q_{\Lambda}$.
We will assume that dark energy is of the vacuum energy type $\rho_{\Lambda}=-p_{\Lambda}$, and as  mentioned above, energy conservation requires $Q_{\Lambda}=-Q_{re}$. Since the dark energy fluid dominates the energy density of the Universe, we have $\rho_{\Lambda}\approx (3/8\pi) H^2m_p^2$. If we write $Q_{\Lambda}=-\Gamma\rho_{\Lambda}$, then we have $\Gamma =  1/(3\pi)H \Lambda^2/m_p^2$. The continuity equation implies that $\dot\rho_{\Lambda} = -\Gamma\rho_{\Lambda}$. It is useful to compare with standard continuity equation for a fluid with equation of state $w$, $\dot\rho+3H(1+w)\rho=0$. We find that the effective equation of state of dark energy, when taking into account the source term, becomes
\beq \label{weff}
w_{eff} =  \frac{1}{9\pi}\frac{\Lambda^2}{m_p^2}-1~.
\eeq
If we take the cutoff to be the fundamental Planck scale, $\Lambda = m_p$, then we obtain $w_{eff}\approx  -0.96$. 

We could also have derived the result of eq.~(\ref{weff}) in a simpler but less illuminating fashion by just using $p=w_{eff} \rho$, with $\rho= \rho_{\Lambda} +\rho_{re}$ and $p= p_{\Lambda} +p_{re}$

\subsubsection*{Note added in proof} After this manuscript was submitted to the essay competition of the Gravity Research Foundation on March 29, 2010, two papers have subsequently appeared with similar observations and conclusions \cite{Maggiore:2010wr,Mangano:2010hw}.





\begin{thebibliography}{99}

\bibitem{Riess:1998cb}
  A.~G.~Riess {\it et al.}  [Supernova Search Team Collaboration],
  Astron.\ J.\  {\bf 116} (1998) 1009
  [arXiv:astro-ph/9805201].

\bibitem{Astier:2005qq}
  P.~Astier {\it et al.}  [The SNLS Collaboration],
  Astron.\ Astrophys.\  {\bf 447} (2006) 31
  [arXiv:astro-ph/0510447].

\bibitem{WoodVasey:2007jb}
  W.~M.~Wood-Vasey {\it et al.}  [ESSENCE Collaboration],
  Astrophys.\ J.\  {\bf 666} (2007) 694
  [arXiv:astro-ph/0701041].

\bibitem{Spergel:2006hy}
  D.~N.~Spergel {\it et al.}  [WMAP Collaboration],
  Astrophys.\ J.\ Suppl.\  {\bf 170} (2007) 377
  [arXiv:astro-ph/0603449].

\bibitem{Tegmark:2006az}
  M.~Tegmark {\it et al.}  [SDSS Collaboration],
  Phys.\ Rev.\  D {\bf 74} (2006) 123507
  [arXiv:astro-ph/0608632].



\bibitem{Wetterich:1987fm}
  C.~Wetterich,
  Nucl.\ Phys.\  B {\bf 302} (1988) 668.

\bibitem{Peebles:1987ek}
  P.~J.~E.~Peebles and B.~Ratra,
  Astrophys.\ J.\  {\bf 325} (1988) L17.

\bibitem{Ratra:1987rm}
  B.~Ratra and P.~J.~E.~Peebles,
  Phys.\ Rev.\  D {\bf 37} (1988) 3406.

\bibitem{Caldwell:1997ii}
  R.~R.~Caldwell, R.~Dave and P.~J.~Steinhardt,
  Phys.\ Rev.\ Lett.\  {\bf 80} (1998) 1582
  [arXiv:astro-ph/9708069].



\bibitem{KeskiVakkuri:2003vj}
  E.~Keski-Vakkuri and M.~S.~Sloth,
  JCAP {\bf 0308} (2003) 001
  [arXiv:hep-th/0306070].


\bibitem{Polarski:1995jg}
  D.~Polarski and A.~A.~Starobinsky,
  Class.\ Quant.\ Grav.\  {\bf 13} (1996) 377
  [arXiv:gr-qc/9504030].


\bibitem{Tanaka:2000jw}
  T.~Tanaka,
  arXiv:astro-ph/0012431.


\bibitem{Starobinsky:2001kn}
  A.~A.~Starobinsky,
  Pisma Zh.\ Eksp.\ Teor.\ Fiz.\  {\bf 73} (2001) 415
  [JETP Lett.\  {\bf 73} (2001) 371]
  [arXiv:astro-ph/0104043].

\bibitem{Maggiore:2010wr}
  M.~Maggiore,
  arXiv:1004.1782 [astro-ph.CO].

\bibitem{Mangano:2010hw}
  G.~Mangano,
  Phys.\ Rev.\  D {\bf 82} (2010) 043519
  [arXiv:1005.2758 [astro-ph.CO]].








\end{thebibliography}
\end{document}